\documentclass[conference]{IEEEtran}

\makeatletter
\def\ps@headings{%
\def\@oddhead{\mbox{}\scriptsize\rightmark \hfil \thepage}%
\def\@evenhead{\scriptsize\thepage \hfil \leftmark\mbox{}}%
\def\@oddfoot{}%
\def\@evenfoot{}}
\makeatother
\usepackage{caption}
\usepackage{fancyhdr}
\usepackage[dvips]{graphicx}
\usepackage{listings}
\usepackage{amssymb}
\usepackage{amsmath}
\usepackage{amsthm}
\usepackage{color}
\usepackage{multirow}
\usepackage{placeins}
\usepackage{subfig}
\usepackage{epsfig}

\newcommand{\eat}[1]{}

\usepackage{color}

\newcommand{\todo}[1]{}

\newtheorem{theorem}{Theorem}

\newcommand{\ceil}[1]{\left\lceil #1 \right\rceil}

\newcommand{\comment}[1]{}

\begin{document}
\title{Energy-Aware Load Balancing in Content Delivery Networks}
\author{\IEEEauthorblockN{Vimal Mathew$^\dag$, Ramesh K. Sitaraman$^\dag$$^\ddag$ and Prashant Shenoy$^\dag$}
\IEEEauthorblockA{$^\dag$University of Massachusetts, Amherst\hspace{0.2in}$^\ddag$Akamai Technologies Inc.}
}
\maketitle


\begin{abstract}
  Internet-scale distributed systems such as content delivery networks
  (CDNs) operate hundreds of thousands of servers deployed in
  thousands of data center locations around the globe. Since the
  energy costs of operating such a large IT infrastructure are a
  significant fraction of the total operating costs, we  argue for redesigning CDNs to incorporate energy optimizations as a
  first-order principle. We propose techniques to turn off CDN servers
  during periods of low load while seeking to balance 
  three key design goals: maximize energy reduction, 
  minimize  the impact on client-perceived service availability (SLAs), and
  limit the frequency of on-off server transitions to reduce
  wear-and-tear and its impact on hardware reliability. We propose an
  optimal offline algorithm and an online algorithm to extract
  energy savings both at the level of local  load balancing within
  a data center and global load balancing across data centers. We
  evaluate our algorithms using real production workload traces from a
  large commercial CDN. Our results show that it is possible to
  reduce the energy consumption of a CDN by more than $55\%$ while ensuring 
  a high level of availability that meets customer SLA requirements and incurring an average of  one on-off transition per server per day. Further, we show that keeping even $10\%$ of the servers as hot spares helps absorb load spikes due to global flash crowds with little impact on availability SLAs. 
Finally, we show that redistributing load across proximal data centers can enhance service availability significantly, but has only a modest impact on energy savings.
  \end{abstract}

\section{Introduction}
\label{sec:intro}

Large Internet-scale distributed systems deploy hundreds of thousands
of servers in thousands of data centers around the world.
Such systems currently provide the core distributed infrastructure for
many popular Internet applications that drive business, e-commerce,
entertainment, news, and social networking. The energy cost of
operating an Internet-scale system is already a significant fraction
of the total cost of ownership (TCO) \cite{barroso2007case}.
The environmental
implications are equally profound. A large distributed platform with
100,000 servers will expend roughly 190,000 MWH per year, enough
energy to sustain more than 10,000 households. In 2005, the total data
center power consumption was already 1\% of the total US power
consumption while causing as much emissions as a mid-sized nation such
as Argentina. Further, with the deployment of new services and the
rapid growth of the Internet, the energy consumption of
data centers is expected to grow at a rapid pace of more than $15\%$
per year in the foreseeable future \cite{koomey2008}. These factors
necessitate a complete rethinking of the fundamental architecture of
Internet-scale  systems to include energy optimization
as a first-order principle.
 
An important Internet-scale distributed system to have evolved in the past
decade is the content delivery network (CDN, for short) that delivers
web content, web and IP-based applications, downloads, and streaming
media to end-users (i.e., {\em clients}) around the world
\cite{dilley2002globally}. A large CDN, such as that of a commercial
provider like Akamai,
consists of hundreds of thousands of servers located in over a thousand
data centers around the world and account for a significant fraction
of the world's enterprise-quality web and streaming media traffic
today \cite{nygren2010akamai}. The servers of a CDN are deployed in
{\it clusters\/} where each cluster consists of servers in a
particular data center in a specific geographic location. The clusters
are typically widely deployed on the ``edges''  of the Internet in most major
geographies and ISPs around the world 
so as to be proximal to clients. Clusters can vary in size from tens of
servers in a small Tier-3 ISP to thousands of servers in a large Tier-1
ISP in a major metro area. A CDN's servers cooperatively deliver
content and applications to optimize the {\em availability\/} and {\em
  performance\/} experienced by the clients. Specifically, each client
request is routed by the CDN's {\em load balancing system\/} to an
``optimal'' server that can serve the content with high availability
and performance. Content and applications can typically be replicated
on demand to any server of the CDN. The load balancing system ensures
high availability by routing each client request to an appropriate
server that is both  {\em live} and {\em not overloaded}. Further, the load
balancing system ensures good performance by routing each client
request to a cluster that is {\em proximal} to that client. For
instance, a client from a given metro area would be routed to a server
cluster in the same metro area or perhaps even the same last-mile
network. The proximity (in a network sense) of the client and the
server ensures a communication path with low latency and loss. A
comprehensive discussion of the rationale and system architecture of
CDNs is available in \cite{nygren2010akamai}.

\noindent{\bf Problem Description.}
In this paper,  we focus on reducing the energy consumption of large
Internet-scale distributed systems, specifically CDNs. Energy
reduction in CDNs is a multi-faceted problem requiring advances in the
power usage effectiveness (PUE) of data centers, improvements in
server hardware to make them more ``energy proportional''
\cite{barroso2007case}, as well as advances in the architecture of CDN
itself.   Our focus is on the CDN architecture, and more specifically,
on its load balancing system. Recent work in server energy management
has suggested the technique of utilizing deep-sleep power-saving modes or even completely turning off servers during periods of low load, thereby saving the energy expended by idle servers \cite{chen2008energy, lin2011dynamic}. We explore the potential applicability of this technique in the CDN context where it is important to understand the interplay of the three objectives below.
\begin{list}{\labelitemi}{\leftmargin=0.5em}
\item {\em  Maximize energy reduction. \/}  Idle servers often consume more than 50\% of
  the power of a fully-loaded one \cite{barroso2007case}. This provides the
  opportunity to save energy by ``rebalancing'' (i.e., redirecting)
  the  request traffic onto fewer servers and turning the remaining servers off. 
\item {\em Satisfy customer SLAs.} Content providers who are the CDN's
  customers would like their content and applications to be served
  with a high level of availability and performance to their clients.
  Availability can be measured as the fraction of client requests that
  are successfully served. A typical SLA would require at least ``four
  nines'' of {\em end-to-end} availability (i.e., 99.99\%).     To
  achieve this end-to-end SLA goal, 
  we estimate that any acceptable technique for powering off servers
  should cause no more than a loss of $0.1$ basis points of
  availability in the data center, leading us to target $99.999\%$
  {\em server} availability with our techniques.  
  In addition to the
  availability SLA, the content providers also require good
  performance. For instance, clients downloading http content should
  experience small download times and clients watching media should
  receive high quality streams with high bandwidth and few freezes.
 Since turning off servers to save energy reduces the live server capacity used for serving the incoming request load, it is
 important that any energy saving technique minimizes the
 impact of the decreased capacity on  availability and 
 performance.  
\item {\em Minimize server transitions.}   Studies have shown that
  frequently turning an electronic device on and off  can impact its
  overall lifetime and reliability. Consequently, CDN
operators are often concerned about the wear and tear caused by
excessive on-off server transitions  that could potentially decrease
the lifetime of the servers.
Additionally, when a server is turned off, its  state has to be
migrated  or replicated to a different live server. Mechanisms for
replicating content footprint and migrating long-standing TCP
connections exist in the CDNs today \cite{nygren2010akamai} as well as
in other types of Internet-scale services
\cite{amur2010robust,chen2008energy}. However, a  small degree of
client-visible performance degradation due to server transitions is
inevitable.   Consequently an energy saving technique should 
limit on-off server transitions in order to reduce 
wear and tear and the impact on client-visible performance.
\end{list}
The three objectives above are often in conflict. For instance,
turning off too many servers to maximize energy reduction can decrease
the available live capacity of the CDN. Since it takes time to turn on a
server and bring it back into service, an unexpected spike in the load
can lead to dropped requests and SLA violations. Likewise, turning
servers on and off frequently in response to load variations could enhance energy reduction but incur
too many server transitions. Our goal is to design energy-aware
techniques for CDNs that incorporate all three objectives and to
understand how much energy reduction is realistically achievable in
a CDN. Since CDNs are yet to be aggressively optimized for energy usage
today, our work hopes to  guide the future architectural
evolution that must inevitably incorporate energy as a primary design
objective.

While we focus on CDNs, our work also applies to
other CDN-like distributed systems that replicate services within and
across server clusters and employ some form of load balancing to
dynamically route requests to servers. On a different dimension, it is
also important to note that our focus is energy {\it usage\/}
reduction rather than energy {\it cost\/} reduction. Note that energy
cost reduction can be achieved by dynamically shifting the server load
to locations with lower energy prices without necessarily decreasing
the total energy usage \cite{qureshi2009cutting}.

\noindent{\bf Research Contributions.}
Our work is the first to propose energy-aware mechanisms for load balancing in CDNs  with a quantification of the key practical tradeoffs between energy reduction, hardware wear-and-tear due to server transitions, and service availability that impacts customer SLAs. The load balancing system of a CDN operates at two
levels \cite{nygren2010akamai}.  The {\em global load balancing\/} component determines a good cluster
of the CDN for each request, while the {\em local load balancing\/}
component chooses the right server for the request within the assigned
cluster. We design mechanisms for energy savings, both
from the local and global load-balancing standpoint. Further, we evaluate
our mechanisms using real production workload
traces collected over 25 days from 22 geographically distributed clusters across the US from a large commercial CDN. Our specific key contributions are as follows.
\begin{list}{\labelitemi}{\leftmargin=0.5em}
\item In the offline context when the complete load sequence for a cluster is known ahead of time, we derive optimal algorithms that minimize energy usage by  varying  the number of live servers required to serve the incoming load.
\item On production CDN workloads, our offline algorithm achieves a
  significant system-wide energy reduction of $64.2\%$.  Further, even
  if the average transitions  is restricted to be below $1$ transition
  per server per day, an energy reduction of $55.9\%$ can be achieved,
  i.e., $87\%$ of the maximum energy reduction can be achieved with
  minimal server wear-and-tear.

\item We propose a load balancing algorithm called $\mathrm{Hibernate}$ that works in an online fashion that makes decisions based on past and current load but not future load, much like a real-life load balancing system.  $\mathrm{Hibernate}$ achieves an energy reduction of $60\%$, i.e.,  within  $94\%$ of the offline optimal.

\item By holding an extra $10\%$ of the servers as live spares,
  $\mathrm{Hibernate}$ achieves the sweet spot with respect to all
  three metrics. Specifically, the algorithm achieves a system-wide
  energy reduction of $55\%$ and a service availability of at least
  five nine's ($99.999\%$),  while incurring an average of at most $1$
  transition per server per day. The modest decrease in energy
  reduction due to the extra pool of live servers is well worth the
  enhanced service availability  for the CDN.

\item In a global flash crowd scenario when the load spikes suddenly across all clusters of the CDN, $\mathrm{Hibernate}$ is still able to provide five nine's of service availability and maintain customer SLAs as long as the rate at which load increases is commensurate with the percentage of server capacity that the algorithm keeps as live spares.

\item Energy-aware global load balancing can redistribute traffic across clusters but had only a limited impact on energy reduction. Since load can only be redistributed between proximal clusters for reasons of client performance, these clusters had load patterns that are similar enough to not entail a large energy benefit from load redistribution. However, a $10\%$ to $25\%$ reduction in server transitions can be achieved by redistributing load across proximal clusters. But, perhaps the key benefit of global load balancing is significantly increased service availability. In our simulations, global load balancing enhanced service availability to almost $100\%$. In situations where an unpredictable increase in load would have exceeded the live capacity of a cluster causing service disruption, our global load balancing spread the load increase to other clusters with available live capacity.
\end{list}
In summary, our results show that significant energy reduction is
possible in CDNs 
if these systems are rearchitected with energy awareness as a first-order principle. Further, our work also allays the two primary fears in the mind of CDN operators regarding turning off servers for energy savings: the ability to maintain service availability, especially in the presence of a flash crowd, and the impact of server transitions on the hardware lifetimes and ultimately the capital expenditures associated with operating the CDN.

\comment{
In the design and evaluation of
our energy-aware mechanisms, we seek to answer the following research
questions: 
\begin{list}
\item  From a local load balancing perspective, what is the maximum energy savings that can be achieved by an optimal (offline)
technique with full knowledge of the future workload? What fraction of
these optimal energy savings can be actually achieved by  an online algorithm that
makes server transition decisions in real-time? 

\item From an SLA perspective, what is the impact of turning off servers on
the traffic (requests) that are dropped due to capacity exhaustion? Is
it possible to achieve four 9s of availability for the SLA  despite
these energy optimizations?

\item How many on-off transitions per server are necessary to achieve the
above energy savings? If the data center operator were to limit the
number of server transitions allowed per day (e.g., no more than one or two
on-off transitions per server per day), how much energy savings can
still be achieved?

\item How agile are our energy-saving mechanisms in terms of turning on servers in
the event of a workload spike (flash crowds)?  What is the impact of
our mechanisms  on the SLA  in the presence of a flash crowd?

\item  From a global load balancing perspective, how does grouping data centers within a city or a geographic regions
enhance the  effectiveness of our mechanisms?
\end{itemize}
 Our theoretical and experimental results show that.
\begin{itemize}
\item Insert key results here
\end{itemize}
}

\noindent{\bf Roadmap.}
First, we review background information (Section~\ref{sec:background}) on load balancing in CDNs.
Next, we study local load balancing (Section~\ref{sec:local}) in an offline setting with the assumption that the entire traffic load pattern is known in advance (Section~\ref{subsec:localoffline}), and then extend it to the more realistic online situation where future traffic is unknown (Section~\ref{subsec:localonline}). Then, we explore the gains to be had by moving traffic between clusters via global load balancing (Section~\ref{sec:global}).  Finally, we discuss related work (Section~\ref{sec:related}) and offer conclusions (Section~\ref{sec:conclusions}).
 
 \section{Background}
 \label{sec:background}
 \noindent {\bf CDN Model.}  Our work assumes a global content
 delivery network (CDN) that comprises a very large number of servers
 that are grouped into thousands of clusters. Each cluster is deployed in a single data center
 and its size can vary from tens
 to many thousands of servers. We assume that incoming requests are
 forwarded to a particular server in a particular cluster by the CDN's load balancing
 algorithm.  Load balancing in a CDN is performed at two levels:
 global load balancing, where a user request is sent to an ``optimum''
 cluster, and local load balancing, where a user request is assigned a
 specific server within the chosen cluster. Load balancing can be
 implemented using many mechanisms such as IP Anycast,
 load balancing switches, or most commonly,   the DNS lookup mechanism \cite{nygren2010akamai}. We do not
 assume any particular mechanism, but we do assume that those mechanisms allow load to
 be arbitrarily re-divided and re-distributed among servers, both
 within a cluster (local) and across clusters (global). This is
 a good assumption for typical web workloads that form a significant
 portion of a CDN's traffic.
 
 \noindent{\bf Energy Model.} Since our goal is to minimize energy usage, we model how servers consume energy as a function of load. Based on our own testing of  typical off-the-shelf server configurations used by CDNs, we use the standard linear model \cite{barroso2007case}  where the power (in Watts)  consumed by a server serving load $\lambda$ is 
 \begin{equation}
 power(\lambda) \stackrel{\Delta}{=}  P_{idle} +  (P_{peak} - P_{idle}) \lambda, \label{eq:linearmodel}
 \end{equation}
 where the load $0 \leq \lambda \leq 1$ is the ratio of the actual load to the peak load, $P_{idle}$  is the power consumed by an idle server, and $P_{peak}$  is the power consumed by the server under peak load. We use typical values of $92$ Watts and $63$ Watts for $P_{peak}$ and $P_{idle}$ respectively.
 Though we use the linear energy model above in all our simulations, our algorithmic results hold for any power function that is convex.

 In addition to the energy consumed by live servers that are serving
 traffic, we also capture the energy consumed by servers that are in
 transition, i.e., either being turned off or tuned on. Servers in
 transition cannot serve load but consume energy; this energy
 consumption is due to a number of steps that the CDN must
 perform during shutdown or startup.  When a server is turned off, the
 load balancing system first stops sending any new traffic to the
 server. Further,  the CDN must wait until existing traffic either dies down or is
 migrated off the server. Additionally, the control responsibilities
 of the server would need to be migrated out by performing leader
 election and other relevant processes. Once the server has been
 completely isolated from the rest of the CDN, it can
 be powered down. When a server is turned on, these same steps are
 executed in the reverse. In both cases, a server transition takes
 several minutes and can be done automatically by the CDN software. To
 capture the energy spent during a transition, we model a fixed amount
 of energy usage of $\alpha$ Joules for each server transition, where
 $\alpha$ typically corresponds to $37$ kilo Joules.

 \noindent{\bf Workload Model.}
 The workload entering the load balancing system is modeled as a discrete sequence $\lambda_t$, $1 \leq t  \leq n$, where $\lambda_t$ is the average load in the $t^{th}$ time slot. We always express load in the normalized unit of actual load divided by peak server capacity.\footnote{For simplicity, we assume that the servers in the CDN are homogenous with identical capacities, though our algorithms and results can be easily extended to the heterogenous case.} Further, we assume that each time slot is $\delta$ seconds long and is large enough for the decisions made by the load balancing algorithm to take effect.  Specifically, in our experiments, we choose a typical $\delta$ value of 300 seconds. 

\noindent{\bf Algorithmic Model for Load Balancing.} 
While a real-life load balancing system is complex \cite{nygren2010akamai}, we model only those aspects of such a system that are critical to  energy usage.
For simplicity, our load balancing algorithms redistribute the incoming load rather than explicitly route incoming requests from clients to servers. The major determinant of energy usage is the number of servers that need to remain live (i.e., turned on) at each time slot to effectively serve the incoming load. The exact manner in which load is distributed to those live servers is less important from an energy standpoint. In fact, in the linear energy model described in Equation~\ref{eq:linearmodel}, the precise manner in which load is distributed to the live servers makes no difference to energy consumption.\footnote{In the more general model where the power function is convex,  distributing the load evenly among the live servers minimizes energy consumption.} In reality, the precise manner in which the load is distributed to the live servers does matter greatly from the perspective of managing footprint and other server state. However, we view this a complementary problem to our own and methods exist in the research literature \cite{amur2010robust,chen2008energy} to tackle some of these issues. 

The local load balancing algorithm of a CDN balances load between live servers of a given cluster. In each time interval $t$, the algorithm distributes the load $\lambda_t$ that is incoming to that cluster. Let $m_t$ denote the number of live servers in the cluster. Servers are typically not loaded to capacity. But rather a {\em target load threshold\/} $\Lambda$, $0 < \Lambda \leq 1,$ is set such that the load balancing algorithm attempts to keep the load on each server of the CDN to no more than the fraction $\Lambda$ of its capacity. Mathematically, if $l_{i,t}$ is the load assigned to live server $i$ at time $t$, then 
$ \sum_{i = 1}^{i = m_t} l_{i,t} = \lambda_t$ and $l_{i, t} \leq \Lambda$,  for $1 \leq i \leq m_t$. In addition to serving the current  load, the load balancing algorithm also decides how many additional servers need to be turned on or off. The changes in the live server count made in time slot $t$ is reflected in  $m_{t+1}$ in the next time slot. 

The global load balancing algorithm works in an analogous fashion and distributes the global incoming  load to the various server clusters. Specifically, the global incoming load is partitioned between the server clusters such that no cluster receives more than a fraction $\Lambda$ of its  capacity. Further,  clients are mapped to proximal clusters to ensure good performance.

\noindent {\bf Online versus Offline.} The load balancing algorithms
work in an {\it online\/} fashion where decisions are made at time $t$
without any knowledge of the future load $\lambda_{t'}$, $t'>
t$. However,  our work also considers the offline scenario where the load balancing algorithm knows the entire load sequence $\lambda_t, 1 \leq t \leq n$ ahead of time and can use that knowledge to make decisions. The offline algorithms provide the theoretically best possible scenario by making future traffic completely predictable. Thus, our provably-optimal offline algorithms provide a key baseline to which realistic online algorithms can be compared.

\noindent{\bf Metric Definitions.} We are interested in the interplay of three metrics: energy reduction, service availability as it relates to customer SLA's, and server transitions. The energy reduction achieved by an algorithm that can turn servers on or off equals the percentage energy saved in comparison to a baseline where all servers remain turned on for the entire period.  Since most CDNs today  are not aggressively optimized for energy, the baseline is representative of the actual energy consumption of such systems. A server cluster that receives more load than the total capacity of its {\em live\/} servers cannot serve that excess load which must be dropped. The client requests that correspond to the dropped load experience a denial of service. The service availability over a time period is computed as $ 100 * (total\ served\ load)/ (total\ input \ load).$   Finally, the server transitions are expressed either as total amount over the time period, or as an average amount expressed as the number of transitions per server per day.

\noindent{ \bf Empirical Data from the Akamai Network.}
 To validate our algorithms and to quantify their benefits in a
 realistic manner, we used 
extensive load traces collected over  $25$ days from a
 large set of Akamai clusters (data centers) in the US.  The $22$ clusters captured in our traces are distributed widely within the US and had 15439 servers in total, i.e., a nontrivial fraction of Akamai's US deployments.  Our load traces account for a peak traffic of 800K requests/second and an aggregate of  950 million requests delivered to clients.
 The traces consist of a snapshot of  total load served by each cluster collected every 5-minute interval from Dec 19th 2008 to January 12th 2009, a time period that includes the busy holiday shopping season for e-commerce traffic (Figure~\ref{fig:fullload}).  
 \begin{figure}[t]
\centering
\includegraphics[width=0.25\textwidth]{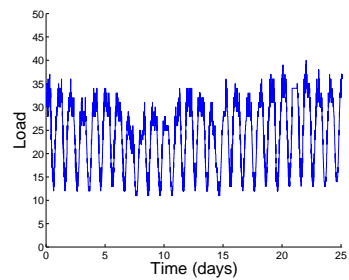}
\caption{Average load per server measured every 5 minutes across $22$
  Akamai clusters in the US over $25$ days. Note load variations due to
  day, night, weekday, weekend, and holidays (such as low load on day
  no. $8$, which was Christmas).}
\label{fig:fullload}
\end{figure}
 
 \section{Local Load Balancing}
 \label{sec:local}

We explore energy-aware algorithms for local load balancing in a CDN. First, we derive optimal offline algorithms that provably provide the maximum energy reduction that is theoretically possible (Section~\ref{subsec:localoffline}). Then, we derive practical online algorithms and evaluate them on realistic load traces from a CDN (Section~\ref{subsec:localonline}), paying particular attention to how well they do in comparison to the theoretical baselines provided by the offline algorithms . 

\subsection{An Optimal Offline Algorithm}
\label{subsec:localoffline}
Given the entire input load sequence, $\lambda_t, 1 \leq t \leq n$,  for a cluster of $M$ servers and a load threshold $\Lambda$, an offline algorithm produces a sequence $m_t$, $1 \leq t \leq n$, where $m_t$ is the number of servers that need to be live at time slot $t$. Note that given the output schedule, it is straightforward to create an on-off schedule for the servers in the cluster to achieve the number of live servers required at each time step. The global load balancing algorithm ensures that the input load sequence can be feasibly served by the cluster if all $M$ servers are live, i.e., $\lambda_t \leq \Lambda M$ for all $1 \leq t \leq n$. In turn, an energy-aware local load balancing algorithm orchestrates the number of live servers  $m_t$ such that  load $\lambda_t$ can be served by $m_t$ servers without exceeding the target load threshold $\Lambda$, i.e., $\lambda_t  \leq \Lambda m_t$, for all $1 \leq t \leq n$.  Assuming that load  $\lambda_t$ is evenly distributed among the  $m_t$ live servers, the  energy expended in the cluster for serving the input load sequence equals
$$\delta \sum_{t = 1}^{t = n} m_t \cdot power(\lambda_t/m_t)  + \alpha \sum_{t = 1}^{t = n}  |m_t - m_{t-1}|,$$
where the first term  is the energy consumption of the live servers and the second is the total  energy  for server transitions. 

We develop an optimal offline local load balancing algorithm $\mathrm{OPT}$  using dynamic programming. Algorithm $\mathrm{OPT}$ produces a  schedule  $m_t, 1 \leq t \leq n,$ that can serve the input load with the smallest energy usage. We construct a two-dimensional table $E(t,m)$ that denotes the minimum energy required to serve the load sequence $\lambda_1, \lambda_2, \cdots, \lambda_t$ while ending with $m$ live servers at time $t$. We assume that the algorithm begins at time zero with all $M$ servers in live state. That is,  $E(0, m) = 0$, if $m = M$, and $E(0, m) = +\infty$, if $m \neq M$. We inductively compute all the entries in the table using the following formula:
\begin{eqnarray}
E(t,m) & = & \min_{0 \leq m' \leq M} \{ E(t-1, m') + \delta m \cdot power(\lambda_t/m) \nonumber \\
 &+ & \alpha \cdot |m - m'|  \}, \  if \ \lambda_t \leq \Lambda m \label{eq:localfeasible} \\
& = & +\infty,\  otherwise \nonumber
\end{eqnarray}
Specifically, if it is feasible to serve the current load $\lambda_t$
with $m$ servers, we extend the optimal solution for the first $t - 1$
steps to the $t^{th}$ step using Equation~\ref{eq:localfeasible}. The
first term in Equation~\ref{eq:localfeasible} is the cost of a
previously computed optimal solution for the first $t - 1$ steps, the
second term denotes the energy consumed by the live servers in time
slot $t$, and the third term denotes the energy consumed in
transitioning servers at time slot $t$. If it is infeasible to serve
the current load with $m$ servers, we set the optimal cost $E(t, m)$
to infinity. Once the table is filled, the optimal solution
corresponds to entry  $E(n,m)$ such that $E(n,m) = \min_{0 \leq s \leq
  M} E(n,s)$. The theorem below follows.
\begin{theorem}
Algorithm $\mathrm{OPT}$ produces an optimal load balancing solution with the smallest energy consumption in time $O(nM^2)$ and space $O(nM)$, where $n$ is number of time slots and $M$ is the number of servers in the cluster.
\end{theorem}

Since we are also interested in knowing how much energy reduction is possible if we are only allowed a small bounded number of server transitions, we develop algorithm $\mathrm{OPT(k)}$ that minimizes energy while maintaining the total number of server transitions to be at most $k$. To this end, we use a three-dimensional table $E(t,m,k),$ $0  \leq t \leq n$, $0 \leq m \leq M$, and $0 \leq k \leq K$. (For simplicity, we assume that all entries of $E(t,m,k)$ with arguments outside the allowable range equal $+\infty$.) $E(t,m,k)$ is the optimum energy required to serve the input load sequence $\lambda_1, \lambda_2, \cdots, \lambda_t$ while ending with $m$ live servers at time $t$ and incurring no more than $k$ transitions in total. Since we start with all servers live at time zero, $E(0, m, k) = 0$,  for all $0 \leq k \leq K$, provided $m=M$. And, $E(0, m, k) = +\infty$, for all $0 \leq k \leq K$, if $m \neq M$. The table is filled inductively using the following formula:
\begin{eqnarray}
E(t,m,k) & = &\hspace*{-.15in} \min_{m-k \leq m' \leq m+k} \{ E(t-1, m', k - |m - m'|) \nonumber\\
& + & \delta m \cdot power(\lambda_t/m)\nonumber\\
& + & \alpha \cdot |m - m'|  \}, \  if \ \lambda_t \leq \Lambda m \\
& = & +\infty,\  otherwise \nonumber
\end{eqnarray}
For each $0 \leq k \leq K$, the optimal energy attainable with at most
$k$ transitions is simply $E(n,m,k)$ such that  $E(n,m,k) = \min_{0
  \leq s \leq M} E(n,s,k)$. The theorem follows.
 \begin{theorem}
Algorithm $\mathrm{OPT(k)}$ produces the optimal solution with the least energy and  no more than $k$ total server transitions.  $OPT(k)$ can be computed for all $0 \leq k \leq K$ in time $O(nM^2K)$ and space $O(nmK)$.
\end{theorem}

\noindent {\bf Empirical Results.}
We ran algorithm $\mathrm{OPT}$ with a typical value of the load
threshold ($\Lambda = 75\%$) on our CDN load traces that encompass $22$
geographically distributed clusters of a large CDN over a span of
 $25$ days. 
\begin{figure}[t]
 \centering
 \includegraphics[width=2in]{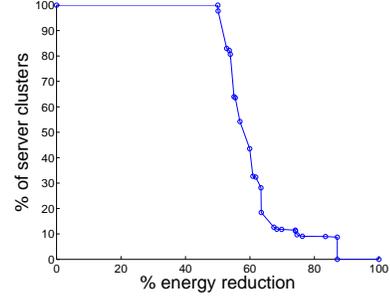}
 \caption{Optimal Offline Energy Reduction. The median cluster achieved a $60\%$ reduction. }
 \label{fig:optoffer}
\end{figure}
\begin{figure}[t]
 \centering
 \includegraphics[width=2in]{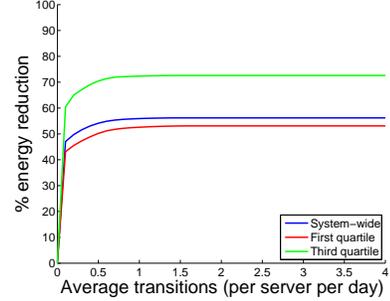}
 \caption{Energy reduction attainable with bounded server transitions. About $87\%$ of the optimal reduction can be achieved with just 1 transition per server per day.}
\label{fig:opttrans}
 \end{figure}
Figure~\ref{fig:optoffer} shows  the percentage of clusters that achieved at least $x\%$ energy reduction, for each value $0 \leq x \leq 100$. For each of the 22 clusters,  $\mathrm{OPT}$  achieved energy reduction in the range $50\%$ to $87\%$. Further, viewing all the clusters of the CDN as a single system, the system-wide energy reduction by using $\mathrm{OPT}$ in all the clusters was $64.2\%$. This implies that significant gains are possible in the offline scenario by optimally orchestrating the number of live servers in each cluster.

Next, we  study how much energy reduction is possible if the server
transitions are  bounded and are required to be
infrequent. Figure~\ref{fig:opttrans} shows the optimal system-wide
energy reduction for each value of the average transitions that is allowable. These numbers
were obtained by running algorithm $\mathrm{OPT(k)}$ for all clusters
for a range of values of $k$. As more transitions are allowed, more
energy reduction is possible since there is a greater opportunity to
turn servers on and off in response to load variations. As the transition bound become large the energy reduction
asymptotically reaches the maximum reduction possible for the
unbounded case of $64.2\%$. The key observation however is that even
with a small number of transitions, say 1 transition per server per
day, one can achieve at least $55.9\%$ system-wide energy reduction in the
offline setting. In other words, with an average of just 1 transition per server per day one can obtain more than  $87\%$ of the energy reduction benefit possible with unbounded transitions.  Besides system-wide energy reduction, Figure~\ref{fig:opttrans} also shows the variation in the energy reduction across clusters by plotting the first and third quartile values for each transition bound.

Note that algorithms $\mathrm{OPT}$ and $\mathrm{OPT(k)}$  never drop any load and achieve an SLA of $100\%$ availability, since they are offline algorithms with complete knowledge of the entire load sequence. After computing the entire sequence of live servers, $m_{t}, 1 \leq t \leq n$, an offline algorithm ensures that  $m_t$ live servers are available at time $t$ by transitioning   $|m_{t} - m_{t-1}|$ servers at time $t-1$. 

\subsection{Online Algorithms}
\label{subsec:localonline}
 \begin{figure*}[t]
 \centering
 \subfloat[][\% Energy reduction decreases as spare servers are kept live longer. ]{\label{fig:hib_right_save}\includegraphics[width=0.25\textwidth]{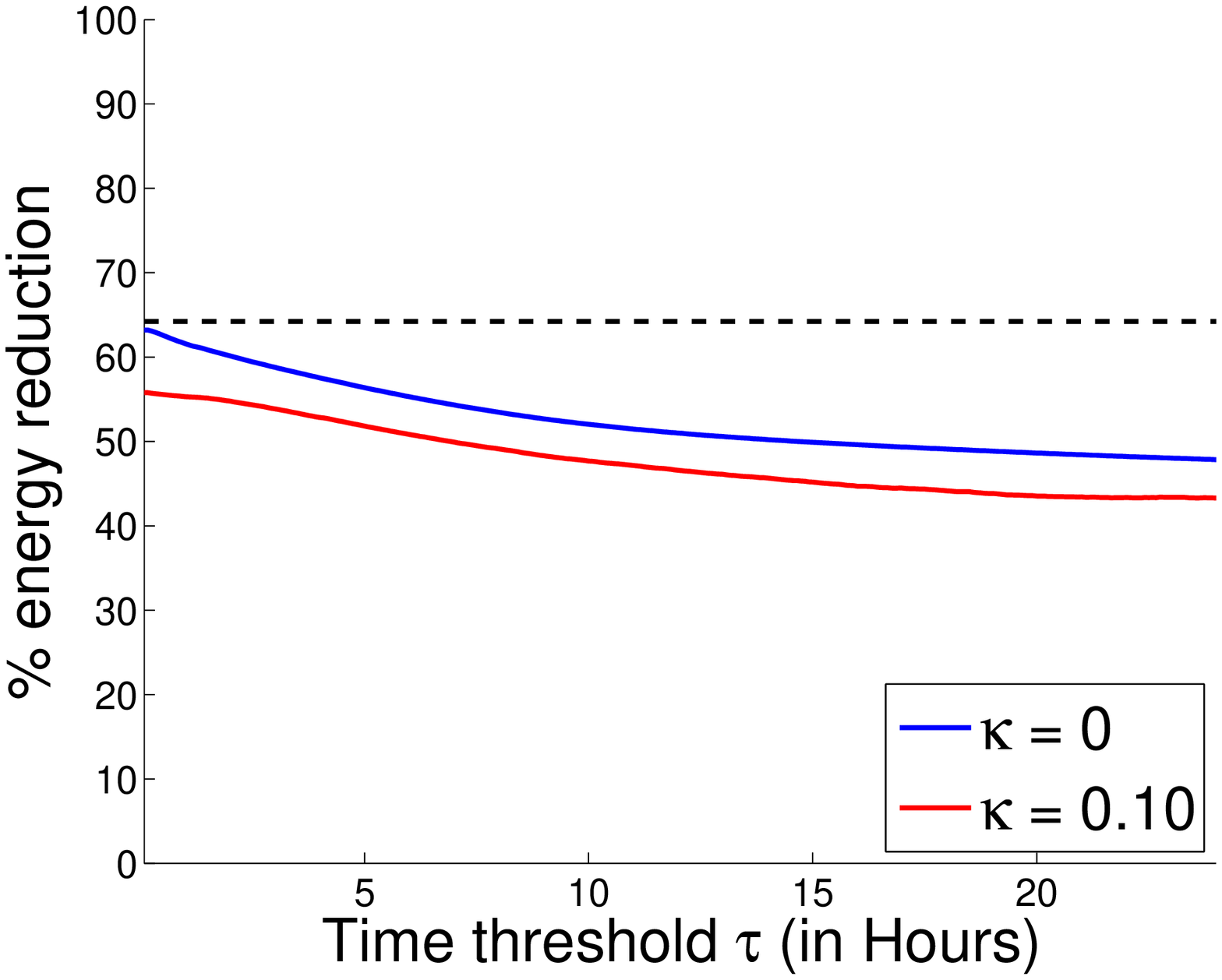}}\hspace*{0.02\textwidth}
 \subfloat[][Average Transitions (per server per day) is small for $\tau = 2$.]{\label{fig:hib_right_trans}\includegraphics[width=0.25\textwidth]{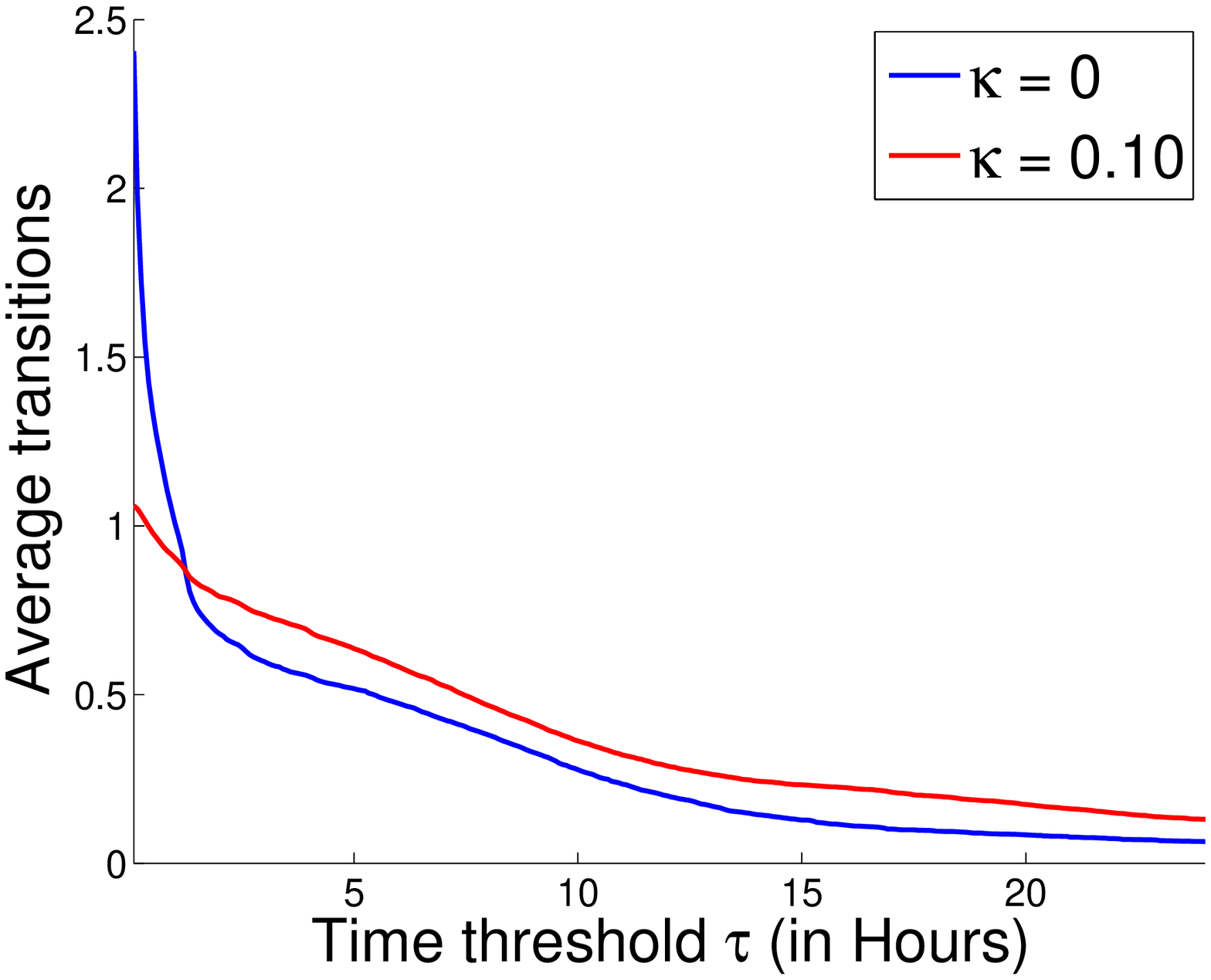}}\hspace*{0.02\textwidth}
 \subfloat[][Availability is significantly enhanced by holding a $10\%$ pool of live spare servers.]{\label{fig:hib_right_barlost}\includegraphics[width=0.25\textwidth]{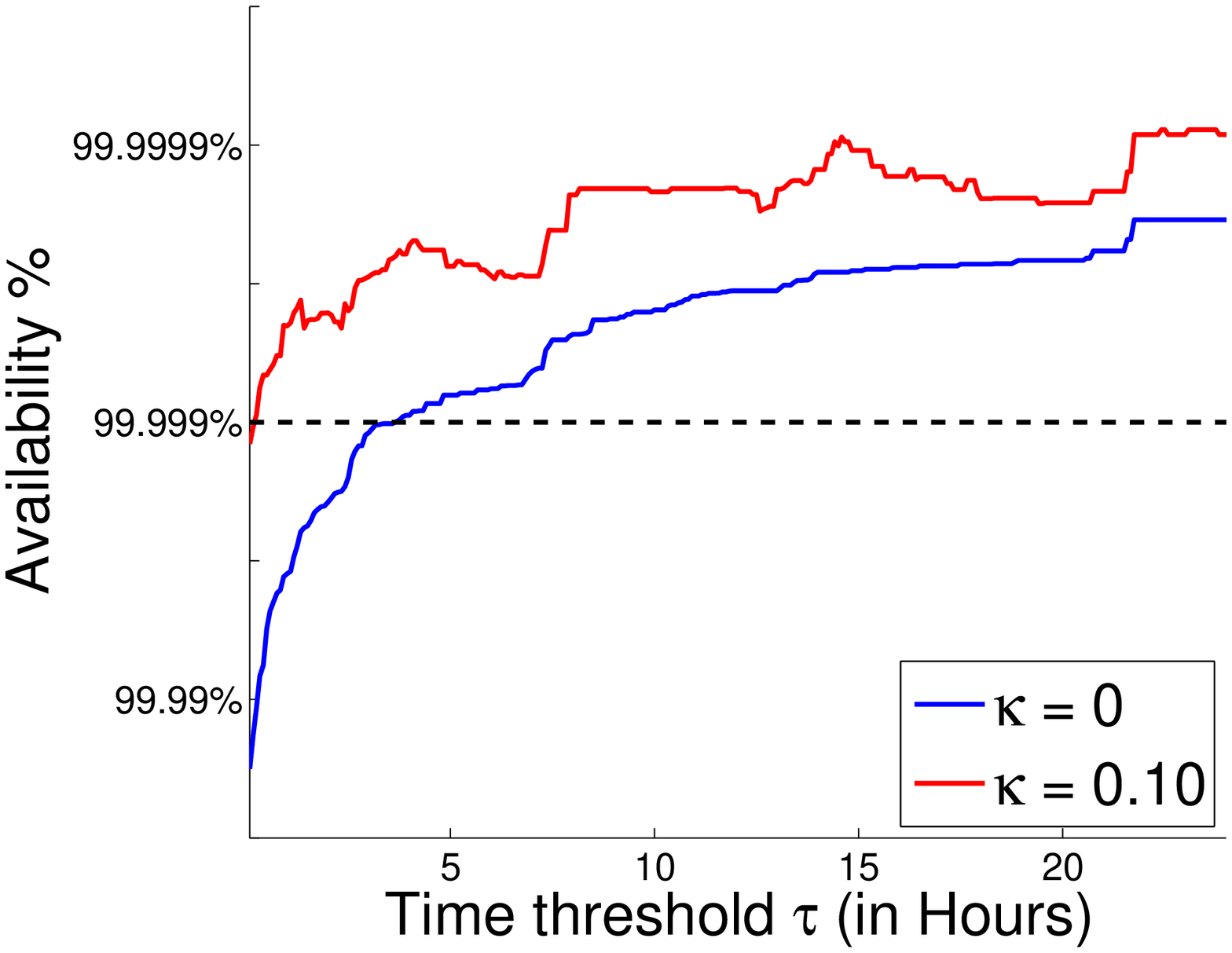}}
 \caption{The three key metrics for algorithm $\mathrm{Hibernate}$ on typical CDN load traces.}
 \label{fig:hibnormal}
\end{figure*}
In contrast to offline algorithms, an online algorithm knows only the
past and current load but has no knowledge of the future load. This
accurately models any real-life load balancing system. At time $t$, an
online algorithm does not know load $\lambda_{t+1}$ and must estimate
the number of servers to transition at the current time step $t$ so
that they are available to serve the load at $t+1$.  Achieving a
balance between the three metrics of energy reduction, transitions,
and service availability that impacts customer SLAs is challenging. If the algorithm keeps a larger number of live
servers to serve future load than is necessary, then the energy
consumption is increased. In contrast, if the algorithm keeps too few
live servers, then some load might have to be dropped leading to
decreased availability and potential customer SLA violations. Our key
contribution in this section is algorithm
$\mathrm{Hibernate}$ that achieves the ``sweet spot'' with respect to
all three metrics, both for typical CDN traffic and flash
crowds. While $\mathrm{Hibernate}$ only uses the past and current load to
make decisions, it is also possible to use workload forecasting
techniques to predict the future workload and use these predictions to
enhance the efficacy of $\mathrm{Hibernate}$. The design of such a
predictive Hibernate is future work.

Algorithm $\mathrm{Hibernate}$  takes two parameters as input, a spare capacity threshold $0 \leq \kappa \leq 1$ and a time threshold $\tau \geq 0$. A key aspect of the algorithm is that it manages a pool of live servers that are considered ``spare'' in the sense that they are in excess of what is necessary to serve the current traffic. Intuitively, spare servers are kept as a  buffer to help absorb unpredictable traffic surges in the future. For simplicity, assume that the servers in the cluster are numbered from $1$ to $M$. Further, assume that the first $m_t$ servers are live  at time $t$, while the rest of the servers are turned off.  At each time $t$, the algorithm does the following.
\begin{list}{\labelitemi}{\leftmargin=0.5em}
\item  Serve the current load $\lambda_t$ using the current set of $m_t$ live servers. If $\lambda_ t > m_t$, the live capacity of the cluster is insufficient to serve the input load. In this case, a load amount of $\lambda_t - m_t$ is dropped and the rest of the load is served.
\item The number of live servers deemed necessary to serve load $\lambda_t$ is  $\ceil{\lambda_t/\Lambda},$ where $\Lambda$ is the target load threshold of the CDN. If $m_t > \ceil{\lambda_t/\Lambda}$, then the live servers numbered $\ceil{\lambda_t/\Lambda} + 1$  to $m_t$ are marked as ``spare''. 

\item The spares are managed according to two rules:
\begin{list}{\labelitemi}{\leftmargin=1em}
\item {\bf Spare Capacity Rule:} Target at least $\ceil{\kappa M}$ servers to be kept as spare, where $0 \leq \kappa \leq 1$. Specifically, if the number of spares $m_t  -\ceil{\lambda_t/\Lambda}$ is smaller than $\ceil{\kappa M}$, then turn on the  $m_t  -\ceil{\lambda_t/\Lambda} - \ceil{\kappa M}$ servers.  (The servers turned on in the current time step t will be live and available to serve load only in the next time step $t+1$.)
\item {\bf Hibernate Rule:} If a server was considered spare in each of the last $\tau$ time slots it is a candidate for being turned off, similar to how a laptop hibernates after a specified period of idleness. However, the hibernate rule is applied only to servers in excess of the spare capacity threshold. Specifically, if the number of spares $m_t  -\ceil{\lambda_t/\Lambda}$ is more than $\ceil{\kappa M}$,  then examine servers  numbered $\ceil{\lambda_t/\Lambda} + \ceil{\kappa M} + 1$ to $m_t$ and turn off any server that was marked as spare in {\em all\/} of the  last $\tau$ time steps.
\end{list}
\end{list}

\noindent{\bf Empirical Results.}
We ran algorithm $\mathrm{Hibernate}$ on typical CDN load traces
collected over 25 days and across 22 clusters for multiple values of
$\tau$ and two values of $\kappa$ with the results summarized in
Figure~\ref{fig:hibnormal}. Note that as the time threshold $\tau$
increases, energy reduction and transitions generally decrease and
availability generally increases. The reason is that as $\tau$
increases, live servers that are spare are turned off after a longer
time period, resulting in  fewer transitions. However, since more
servers are left in an live state, the energy reduction is smaller,
but availability is larger as the additional live servers help absorb
more of the unexpected load spikes. The tradeoff between requiring no
spare capacity ($\kappa = 0$) and requiring a $10\%$ spare capacity
($\kappa = 0.1$) is also particularly interesting. If we fix a typical
value of $\tau = 2$ hours, $\mathrm{Hibernate}$ provides an acceptable
number of transitions ($<1$ transition per server per day) with or
without spare capacity. Requiring 10\% spare capacity decreases the
energy reduction by roughly 10\%, since a pool of spare servers must
be kept live at all times (Figure~\ref{fig:hib_right_save}). However,
the modest decrease in energy reduction may well be worth it for most CDNs, since availability is much higher (five nine's or more) with 10\% spare capacity than with no spare capacity requirement (Figure~\ref{fig:hib_right_barlost}).

\begin{figure*}[t!]
 \centering
\subfloat[][\% Energy reduction increases if the servers  run hotter. The dotted line represents the offline optimal baseline.]{\label{fig:hib_lambda_save}\includegraphics[width=0.26\textwidth]{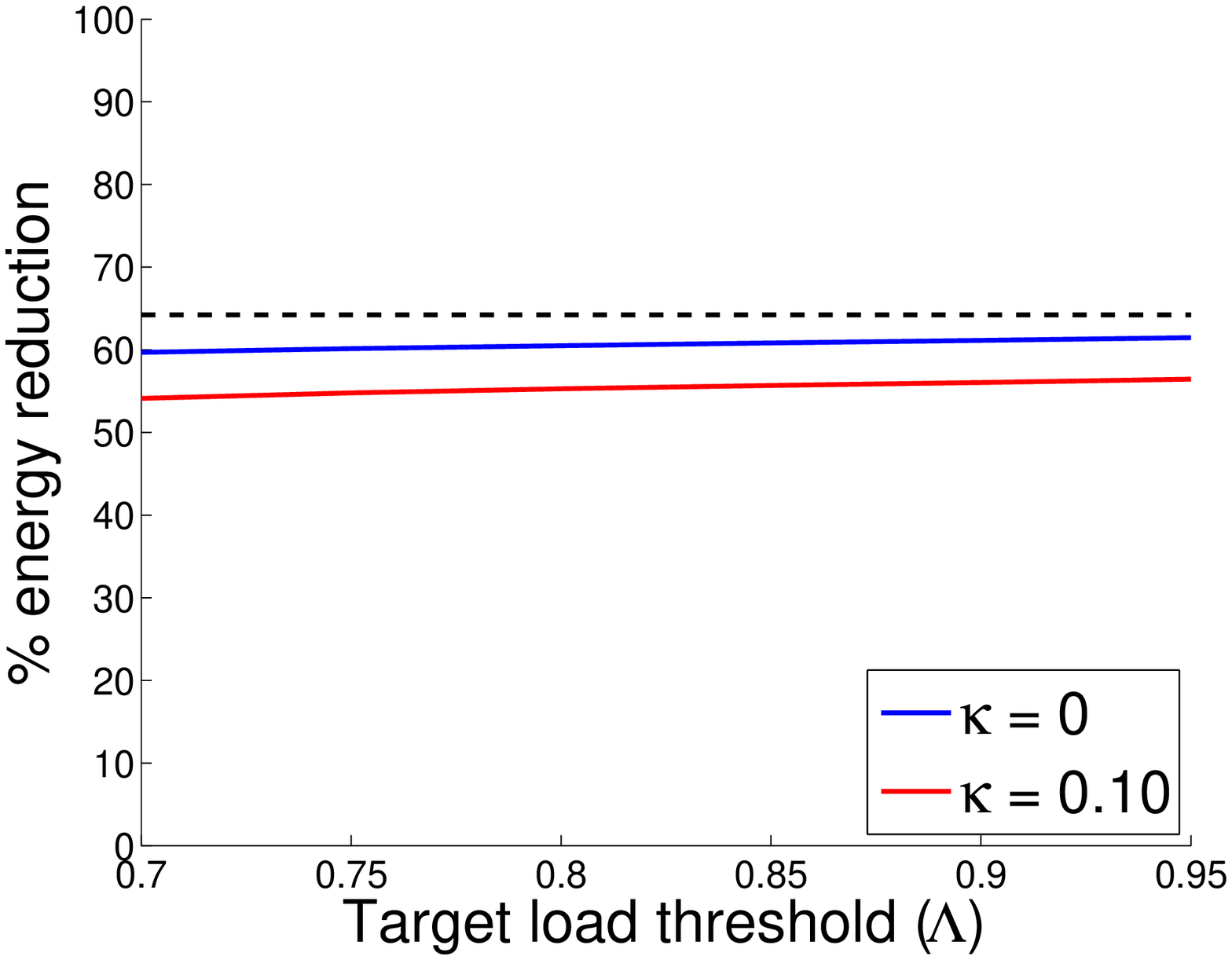}}\hspace*{0.02\textwidth}
 \subfloat[][Average transitions (per server per day) decreases with $\Lambda$ as there is incrementally more capacity in each server.]{\label{fig:hib_lambda_trans}\includegraphics[width=0.26\textwidth]{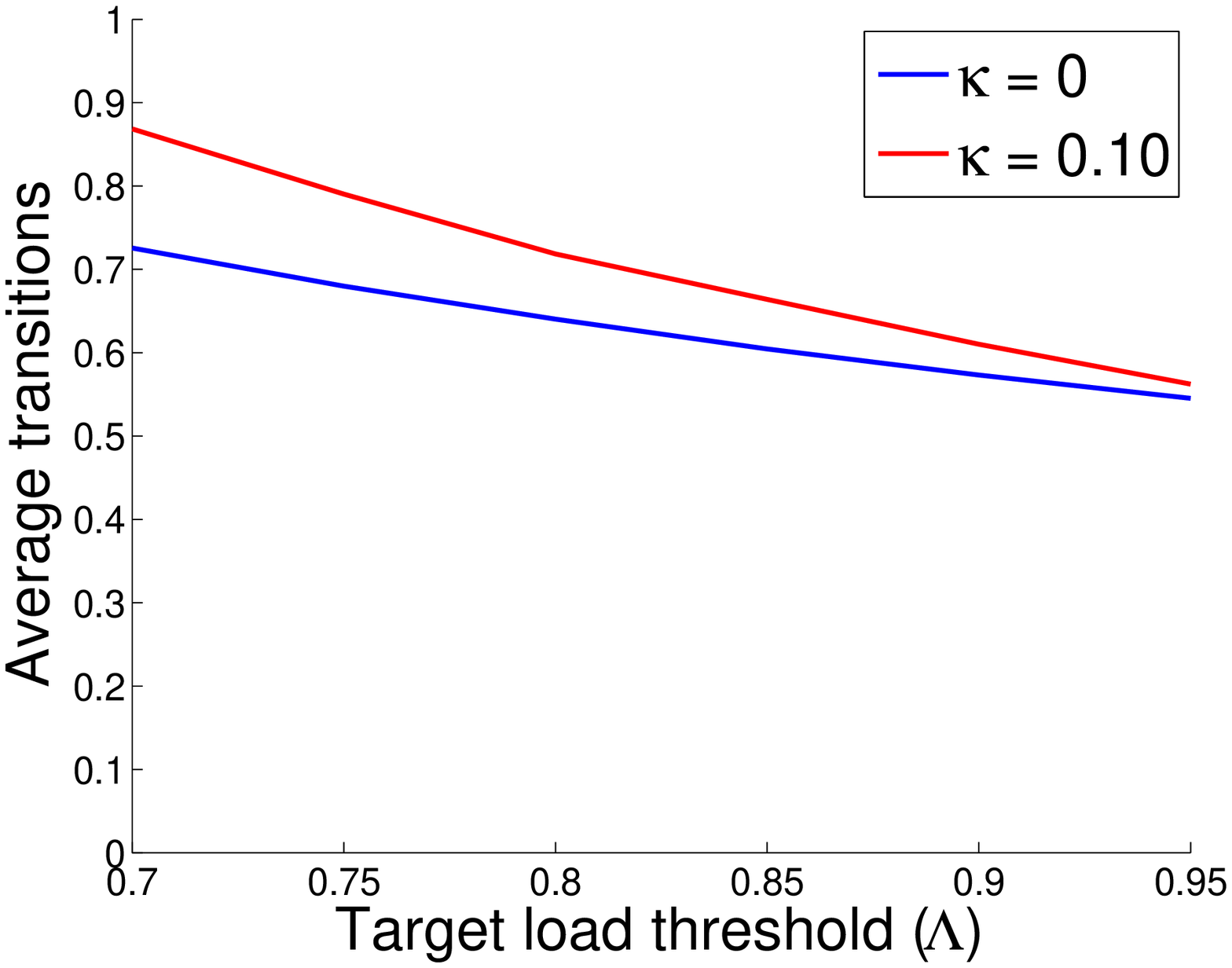}}\hspace*{0.02\textwidth}
\subfloat[][Availability decreases as the system is run hotter but is enhanced by the addition of a $10\%$ pool of live spare servers.]{\label{fig:hib_lambda_avail}\includegraphics[width=0.26\textwidth]{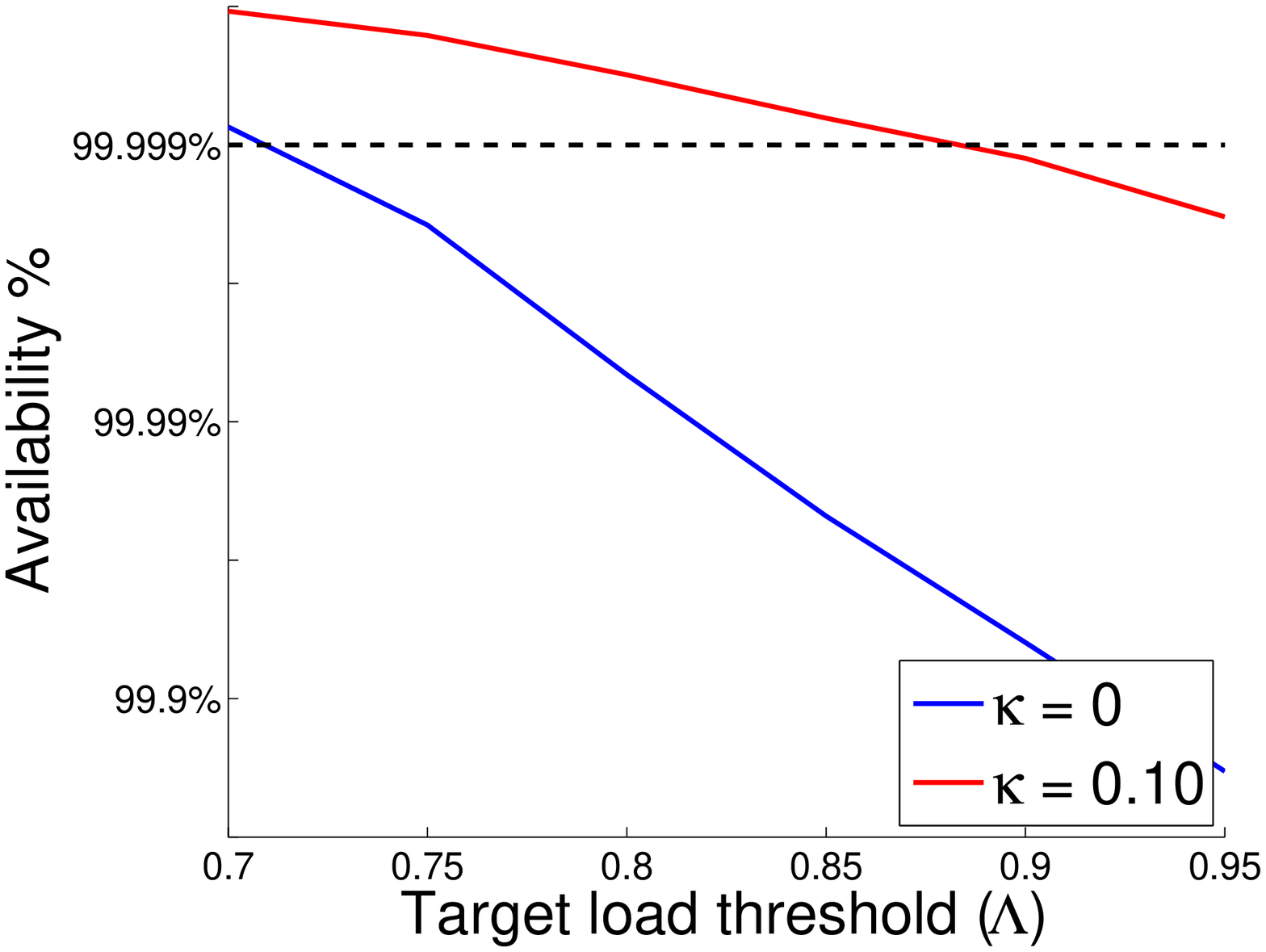}}\\
\caption{Variation of the three metrics with the target load threshold $\Lambda$}
\label{fig:lambda}
\end{figure*}

{\em Handling typical workload fluctuations:} A key decision for a
CDN operator is the target utilization $\Lambda$ that the system
should be run at in order to handle typical workload variations.  The
value of $\Lambda$ is typically kept ``sufficiently'' smaller than $1$
to provide some capacity headroom within each server to account for
the inability to accurately estimate small load variations. In
Figure~\ref{fig:lambda}, we quantify the tradeoffs associated with
$\Lambda$ as it pertains to our three metrics. Running the CDN
``hotter'' by increasing $\Lambda$ would increase the system capacity
and the server utilization. Note that as $\Lambda$ increases, the
effective capacity of each live server increases, resulting in fewer
live servers being needed to serve the load. This results in increased
energy reduction (Figure~\ref{fig:hib_lambda_save}) as well as a
smaller number of transitions
(Figure~\ref{fig:hib_lambda_trans}). However, increasing $\Lambda$ also
decreases availability (Figure~\ref{fig:hib_lambda_avail}) and
potentially increases customer SLA violations. The reason is that by
utilizing the live servers closer to their capacity decreases the
headroom available to buffer temporary load spikes resulting in load
being dropped. Note also that requiring 10\% spares ($\kappa = 0.1$)
allows the CDN operator to run the system hotter with a larger
$\Lambda$ value than if there were no spares ($\kappa = 0$) for the
same availability SLA requirements. Thus, there is a relationship
between the target load/utilization $\Lambda$ and the spares $\kappa$,
since both paramaters permit some capacity ``headroom'' to handle
workload variations. The hotter the system (higher $\Lambda$s), the
more should be the spare threshold $\kappa$ to achieve the same SLA. 

{\em Handling Large Flash Crowds:}
A particular worry of CDN operators from the standpoint of powering
off servers is the global flash crowd scenario where there is a large
unexpected load spike across most clusters of the CDN.  Note that a
local flash crowd scenario that only affects some of the clusters, say
just the northeastern US, is often easier to deal with, since the
global load balancing system will redistribute some of the traffic
outside that local region at some cost to performance.  Global flash
crowds that matter to a large CDN are rare but do occur from time to
time. Some examples include 9/11, and the Obama inauguration. Since it
is critical from the standpoint of a CDN operator to understand the
behavior of any load balancing algorithm in a global flash crowd
situation and since our actual CDN traces lacked a true global flash
crowd event, we modified the traces to simulate one. To pick a
worst-case scenario, we chose a low traffic period in the night when
servers are likely to be turned off and introduced a large spike
measuring $30\%$ percent of the capacity of the cluster and lasting
for a $1$ hour period (Figure~\ref{fig:spike1_sample}.)  Further, to
simulate a global event we introduce the same spike at the same time
in all the 22 clusters distributed across the US. A critical factor in
a flash crowd is the {\em spike rate\/} $\rho$ at which the load
increases (or, decreases) in one time interval (Recall that the time
interval models the ``reaction time'' of the load balancing system
which in our case is 300 seconds). We ran algorithm
$\mathrm{Hibernate}$ for different settings of the spike rate $\rho$
and the spare capacity threshold $\kappa$ with the results summarized
in Figure~\ref{fig:flash}. As $\kappa$ increases, more servers need to
be held live and the energy reduction decreases in a roughly linear
fashion in all the simulated scenarios
(Figure~\ref{fig:spike1_sav}). The average transitions also stayed
within the accepted range of less than 1 transition per server per day
in all cases. However, a direct relationship was observed between the
spike rate $\rho$ and spare capacity threshold $\kappa$ where a larger
$\rho$ was tolerable only with a corresponding larger value of
$\kappa$ to sustain the required levels of service availability and
meeting customer SLAs (Figure~\ref{fig:spike1_lost}). To absorb a
spike rate of $\rho$ with at least five nine's of availability
($99.999\%)$ a commensurately large value of $\kappa$ is required
(Figure~\ref{fig:kappa_rho}). Since the spike rate can be deduced from
prior global flash crowds, this gives clear guidance to CDN
operators on how much spare capacity must be held live at
all times to absorb even large flash crowds.
\begin{figure*}[t!]
 \subfloat[][A simulated  load spike in the \\Ashburn cluster]{\label{fig:spike1_sample}\includegraphics[width=0.24\textwidth]{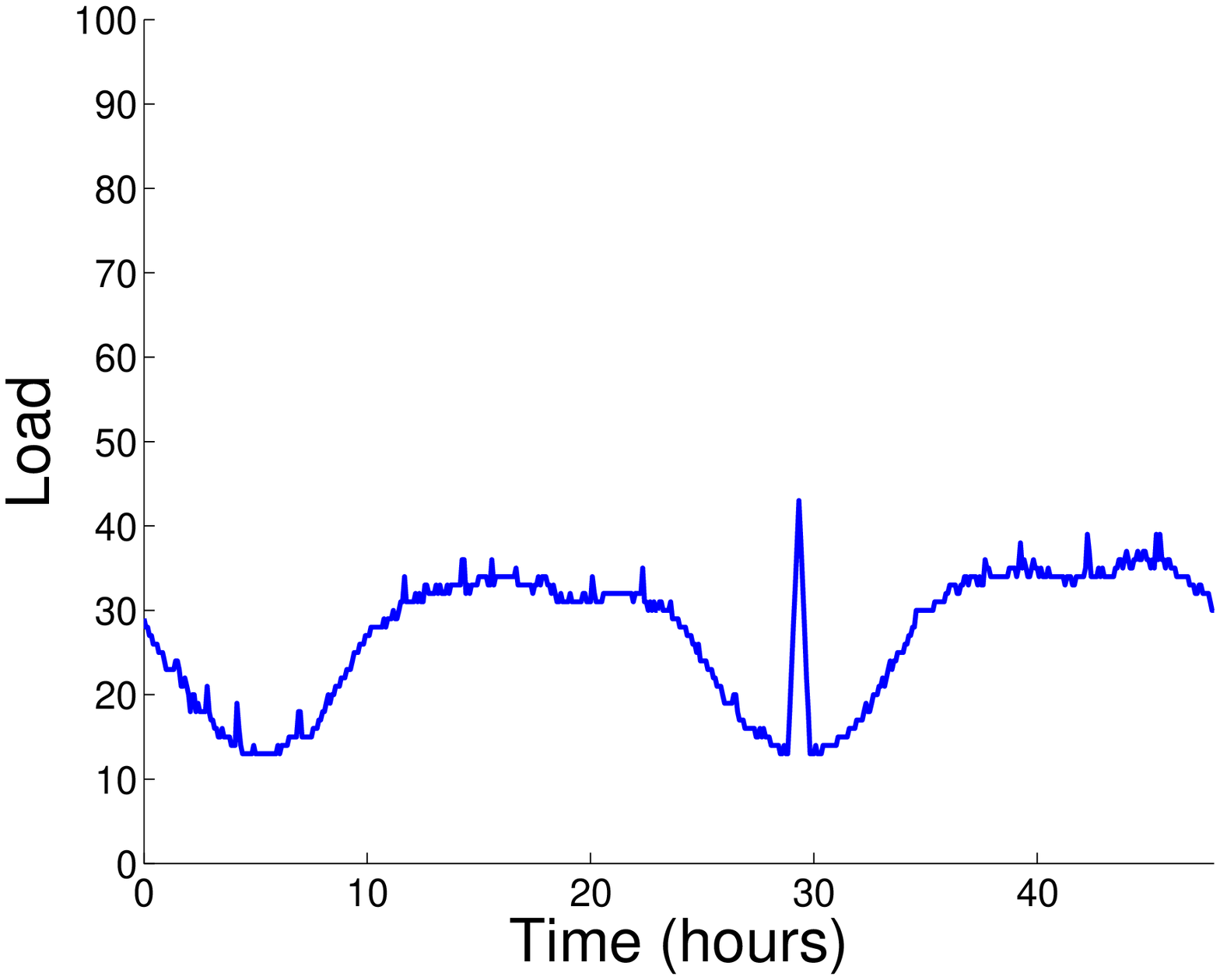}} \subfloat[][\% Energy reduction decreases \\with additional live spares]{\label{fig:spike1_sav}\includegraphics[width=0.24\textwidth]{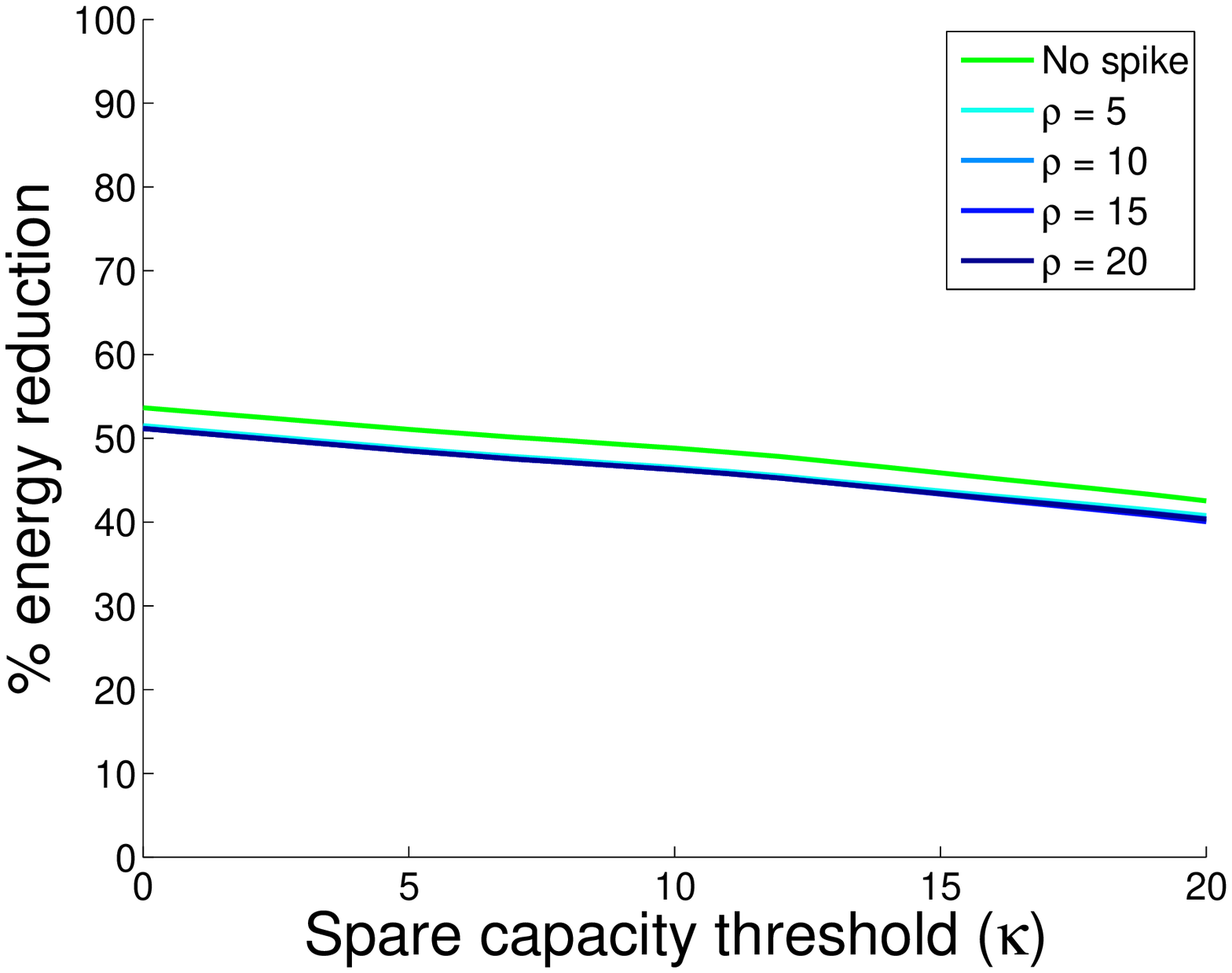}}
\subfloat[][More live spares help\\ increase availability]{\label{fig:spike1_lost}\includegraphics[width=0.24\textwidth]{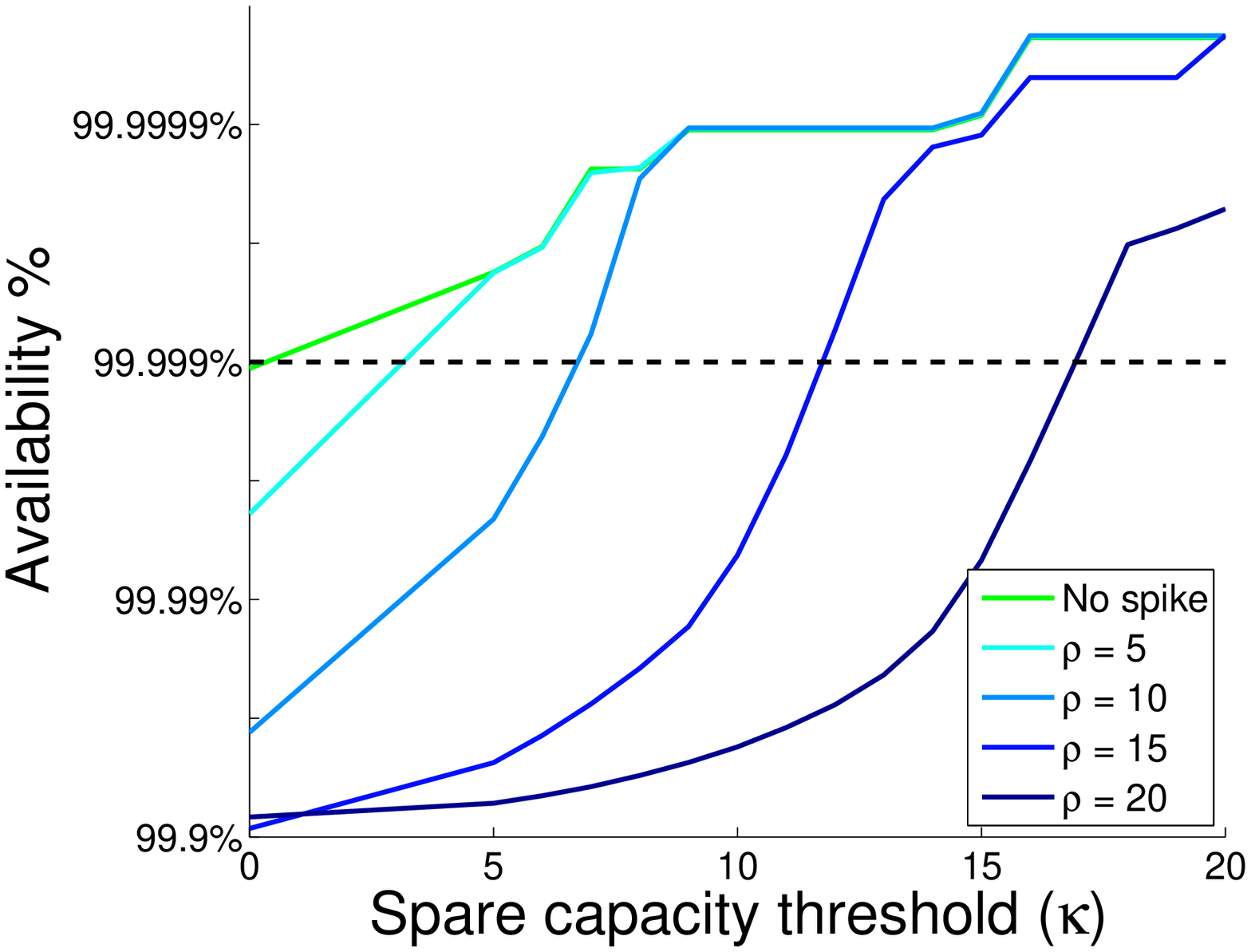}}
\subfloat[][Spares  needed ($\kappa$) to absorb a spike rate $\rho$ with $99.999\%$ availability]{\label{fig:kappa_rho}\includegraphics[width=0.24\textwidth]{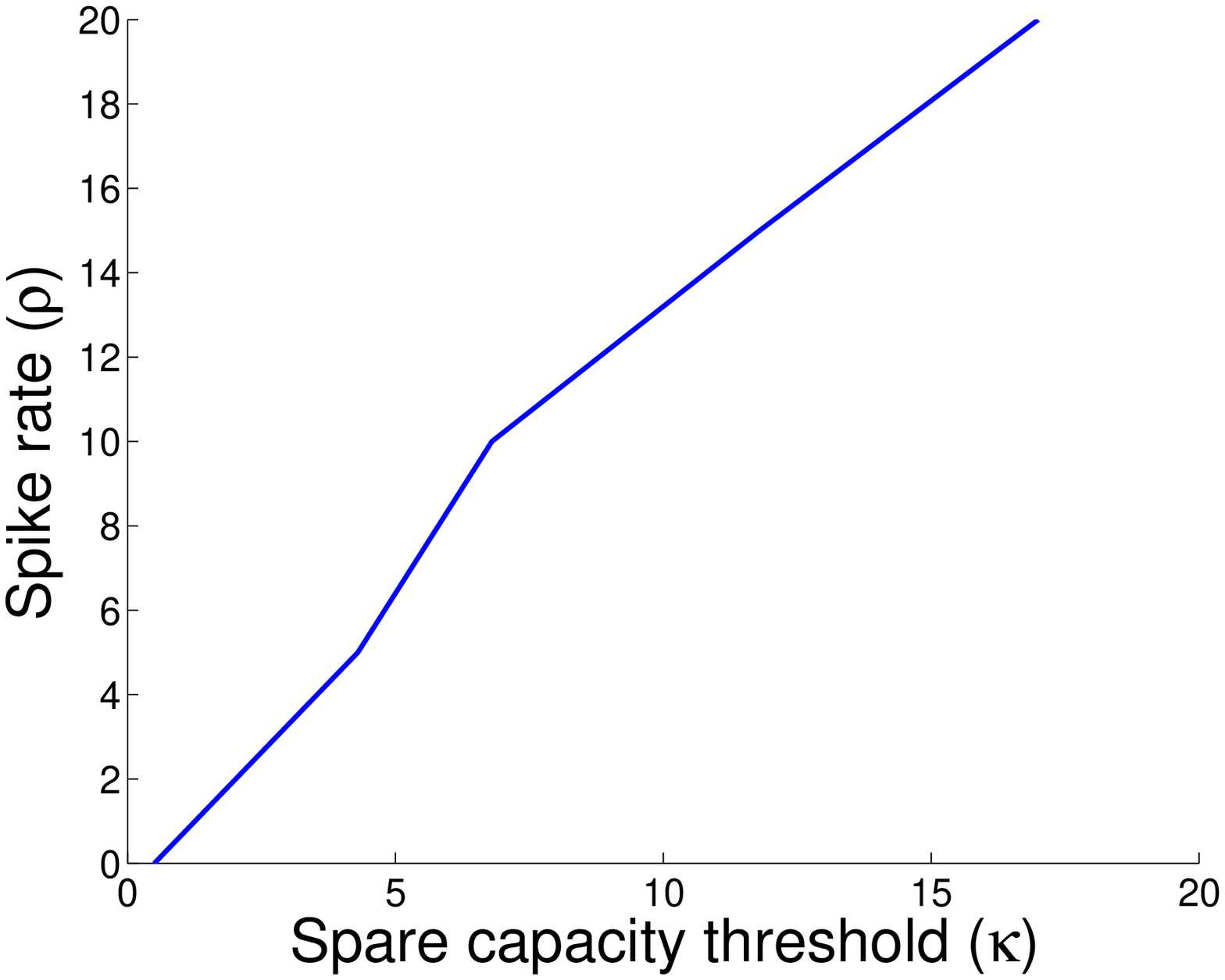}}
\caption{The behavior of $\mathrm{Hibernate}$ during a simulated global flash crowd}
\label{fig:flash}
\end{figure*}

\section{Global Load Balancing}
\label{sec:global}
\begin{figure}[t!]
\centering
\subfloat[][\% Energy reduction]{\label{fig:cluster_red}\includegraphics[width=0.24\textwidth]{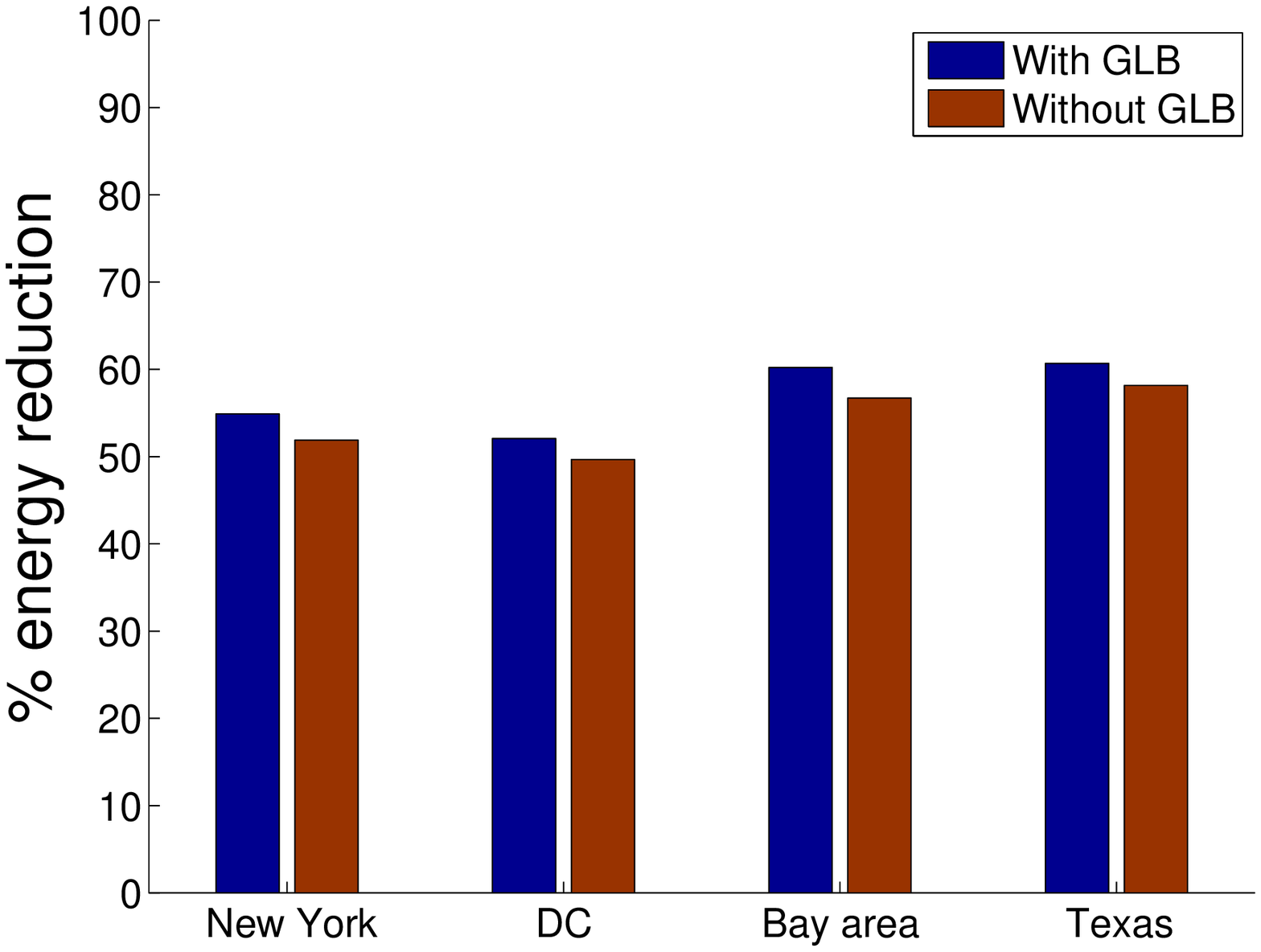}}
\subfloat[][Average Transitions (per server per day)]{\label{fig:cluster_transt}\includegraphics[width=0.24\textwidth]{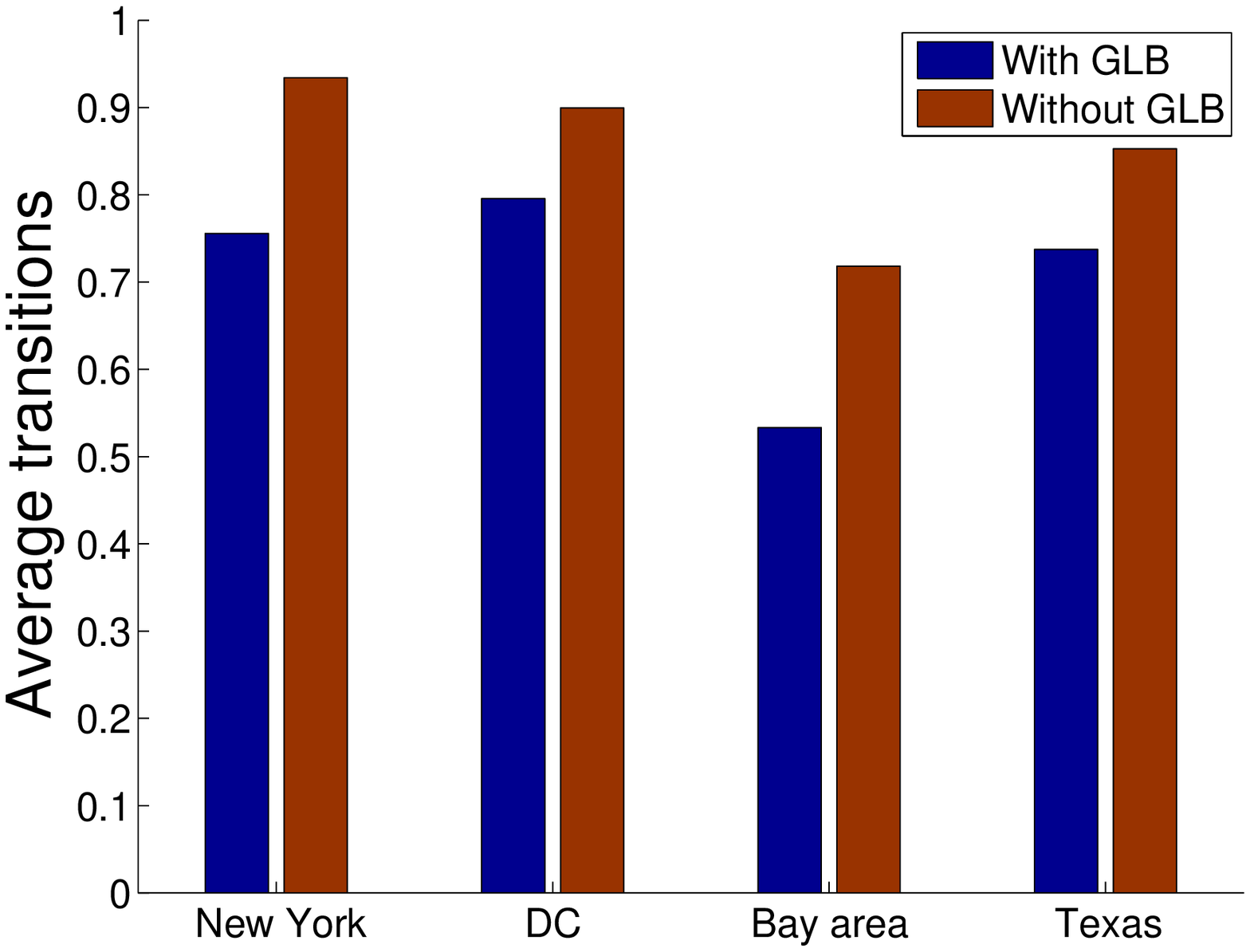}}
\caption{Energy reduction and transitions show only modest improvements with global load balancing}
\label{fig:clusterset}
\end{figure}
\begin{figure}[t!]
\centering
\begin{tabular}{|l|l|l|}
\hline
Cities & With GLB & Without GLB\\
\hline
New York & 100\% & 99.9986\%\\
DC & 100\% & 99.99957\%\\
Bay area & 100\% & 99.9988\%\\
Texas & 100\% & 99.9994 \%\\
\hline
\end{tabular}

\caption{Availability improves drastically with global load balancing}
\label{fig:cluster_avail}
\end{figure}

In prior sections, we devised energy-aware schemes for {\em local\/} load balancing that redistribute  load across servers within\ the {\em same cluster}.  A natural question is what can be gained by energy-aware {\em global\/} load balancing that can redistribute load across  {\em different clusters} of the CDN. An important requirement for global load balancing is that each request is served from a cluster that is ``proximal'' to the client, so as to ensure good network performance. However, a large CDN with wide deployments may have several clusters that can all provide equivalently good performance to a given client. Thus, global load balancing typically has numerous choices of  clusters to serve a given portion of the incoming load. While there are other considerations such as bandwidth costs\cite{AdlerHS2006} that come into play, we focus on energy consumption and ask the following key question. Does redistributing load across clusters that can provide equivalent performance further help optimize energy reduction, transitions, and availability? 

To answer the above  question empirically using our CDN trace data, we create {\em cluster sets} from the 22 clusters for which we have load traces. Each cluster set consists of clusters that are likely to have roughly equivalent performance so as to allow global load balancing to redistribute load between them. To form a cluster set we choose clusters that are located in the same major metropolitan area, since network providers in a major metro area tend to peer well with each other and can likely to provide equivalently good performance to clients from the same area. For instance, our Bay Area cluster set consisted of clusters located in Palo Alto, San Francisco, San Jose, and Sunnyvale, our DC metro area cluster set consists of clusters in Ashburn and Sterling, and our New York metro area cluster set consists of clusters in New York and Newark.  Further, since a large CDN is likely to have more than a dozen clusters in each major metro area and since we only a have trace data for a subset of the clusters of a large CDN, we simulate eight clusters from each actual cluster by dividing up the traces into eight non-overlapping periods of 3-days each and aligning the 3-day traces by the local time of day. To simulate the baseline scenario with no energy-aware global load balancing, we ran our algorithm $\mathrm{Hibernate}$ individually on each cluster. Note that in this case the incoming load to a cluster as represented in the traces is served by the same cluster. Now, to simulate energy-aware global load balancing, we viewed each cluster set as a single large cluster with the sum total of the capacities of  the individual clusters and sum total of the incoming load. We then ran $\mathrm{Hibernate}$ on the large cluster. Note that in this case the incoming load can be redistributed in an arbitrary fashion across the clusters within a cluster set.

The results of our evaluation are summarized in Figures~\ref{fig:clusterset} and \ref{fig:cluster_avail}. The additional energy savings due to global load balancing were modest in the $4\%$ to $6\%$ range. The reason is that clusters within the same cluster set are {\em broadly} similar in the their load patterns, with the peak and off-peak loads almost coinciding in time. Thus, global load balancing is not able to extract significantly more energy savings by moving load across clusters (Figure~\ref{fig:cluster_red}), over and above what can be saved with local load balancing. However, a $10\%$ to $25\%$ reduction in the average transitions can be achieved by global load balancing, since there are occasions where load spikes in one cluster can be served with live spare capacity in a different cluster by redistributing the load rather than incurring server transitions (Figure~\ref{fig:cluster_transt}). But, perhaps the most key benefit of global load balancing is the increased availability (Figure~\ref{fig:cluster_avail}). The enhanced availability is due to an ``averaging'' effect where an unpredictable upward load fluctuation that would have caused some load to be dropped within a single cluster can be routed to a different cluster that happened to have a corresponding downward load fluctuation  leaving some spare live capacity in that cluster. In fact, in our simulations, the availability was nearly $100\%$ with global load balancing in all cluster sets.

\section{Related Work}
\label{sec:related}

Energy management in data centers has been an active area of research
in recent years \cite{ Chen05}. Techniques that have
been developed in this area include, use of DVFS to reduce energy,
use of very low-power servers ~\cite{fawn},
 routing requests to locations with the cheapest energy~\cite{qureshi2009cutting}
and dynamically activating and deactivating nodes as demand rises and
falls~\cite{Chase01,Tolia08,Krioukov10}. A key difference between much
of this prior work and the current work is our focus on CDNs, with a particular emphasis on the
interplay between energy management and the local/global load
balancing algorithms in the CDN.  We also examine the impact of
shutting servers on client SLA as well as  the
impact of server transitions on wear and tear.

A recent effort  related to our work is \cite{lin2011dynamic}. Like us,
this paper also presents offline and online algorithms for turning
servers on and off in data centers. While \cite{lin2011dynamic} targets data
center workloads such as clustered mail servers, our focus is on CDN
workloads. Further, \cite{lin2011dynamic} does not emphasize SLA issues, while
in CDNs, SLAs are the most crucial of the three metrics since violations can result in
revenue losses.  Two recent efforts have considered energy-performance
or energy-QoS tradeoff in server farms \cite{Gandhi10,Kant11}.  Our
empirical results also show an energy-SLA tradeoff,  and we are
primarily concerned with choosing system parameters to obtain five 9s
of availability in CDNs.

\section{Conclusions}
\label{sec:conclusions}

In this paper,
we proposed energy optimization techniques to turn off CDN servers
  during periods of low load while seeking to balance the interplay of
  three key design objectives: maximize energy reduction, 
  minimize  the impact on client-perceived availability (SLAs), and
  limit the frequency of on-off server transitions to reduce
  wear-and-tear and its impact on hardware reliability. We proposed an
  optimal offline algorithm and an online algorithm to extract
  energy savings both at the level of local  load balancing within
  a data center and global load balancing across data centers.  Our
  evaluation using real production workload traces from a
  large commercial CDN showed that it is possible to
  reduce the energy consumption of a CDN by more than $55\%$ while ensuring 
  a high level of availability that meets customer SLA requirements with only a modest number of on-off transitions per server per day. Further, we show that keeping even $10\%$ of the servers as hot spares helps absorb load spikes due to global flash crowds with little impact on availability SLAs. 

Our future work will focus on the incorporation of workload prediction
techniques into our Hibernate algorithm, further optimizations of the
global load balancing algorithm from an energy perspective and
techniques for managing footprint (disk state) of CDN customers while
turning servers on and off.

\section*{Acknowledgment}
The authors would like to acknowledge the support of NSF awards 
CNS-0519894, CNS-0916972, and CNS-1117221. We would like to thank Rick Weber (Akamai) and Bruce Maggs (Duke/Akamai) for their help with data collection. We would also like to thank Tim Dunn (Akamai) and Nicole Peill-Moelter (Akamai) for stimulating discussions about server energy consumption.

\bibliographystyle{plain}
\bibliography{loadbal}

\end{document}